# Parity-Forbidden Transitions and Their Impacts on the Optical Absorption Properties of Lead-Free Metal Halide Perovskites and Double Perovskites

*Weiwei Meng,[†,‡] Xiaoming Wang,[†] Zewen Xiao,[†,⊥] Jianbo Wang,[\*,‡,§] David B. Mitzi,[\*,¶]  and*

*Yanfa Yan[\*,†]*

[†]Department of Physics and Astronomy, and Wright Center for Photovoltaic Innovation and Commercialization, The University of Toledo, Toledo, Ohio 43606, United States

[‡]School of Physics and Technology, Center for Electron Microscopy, MOE Key Laboratory of Artificial Micro- and Nano-structures, and Institute for Advanced Studies, Wuhan University, Wuhan 430072, China

[§]Science and Technology on High Strength Structural Materials Laboratory, Central South University, Changsha 410083, China

[¶]Department of Mechanical Engineering and Materials Science, and Department of Chemistry, Duke University, Durham, NC 27708, United States

ABSTRACT

Using density-functional theory calculations, we analyze the optical absorption properties of lead (Pb)-free metal halide perovskites ($AB^{2+}X_3$) and double perovskites ($AB^+B^{3+}X_6$) (A = Cs or monovalent organic ion, $B^{2+}$ = non-Pb divalent metal, $B^+$ = monovalent metal, $B^{3+}$ = trivalent metal, X = halogen). We show that, if $B^{2+}$ is not Sn or Ge, Pb-free metal halide perovskites exhibit poor optical absorptions because of their indirect bandgap nature. Among the nine possible types of Pb-free metal halide double perovskites, six have direct bandgaps. Of these six types, four show inversion symmetry-induced parity-forbidden or weak transitions between band edges, making them not ideal for thin-film solar cell application. Only one type of Pb-free double perovskite shows optical absorption and electronic properties suitable for solar cell applications, namely those with $B^+$ = In, Tl and $B^{3+}$ = Sb, Bi. Our results provide important insights for designing new metal halide perovskites and double perovskites for optoelectronic applications.

**TOC GRAPHICS**

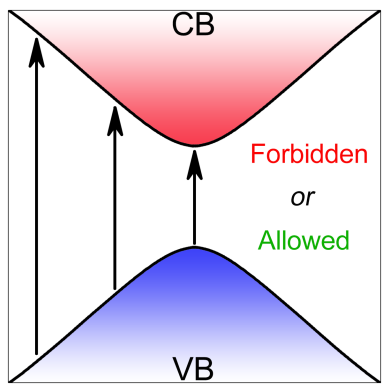

Organic–inorganic lead (Pb) halide perovskite solar cells have attracted significant attention, owing to their rapid improvement in record power conversion efficiency (PCE) over the past few years.[1–6] Despite the demonstration of the great potential, this emerging photovoltaic technology is still facing serious challenges, most notably with regards to cell instability against moisture and temperature and the inclusion of toxic Pb. Extensive efforts have been paid to discover nontoxic or low-toxicity and air-stable metal halide perovskite-based solar cell materials.[7–20]

There are two general routes for designing Pb-free metal halide perovskite related absorber materials. The first route is to replace Pb by a divalent cation to create a Pb-free $AB^{2+}X_3$ perovskite (A = Cs or monovalent organic ion, $B^{2+}$ = non-Pb divalent metal, X = halogen). The divalent cations from group 2, group 12, and group 14 (Ge and Sn) of the periodic table are the choices for the $B^{2+}$ site. Among these possible Pb-free metal halide perovskites, so far, only Sn halides have achieved solar cells with reasonable PCEs.[7–10] Previous studies have revealed the importance of the strong Pb 6s–I 5p antibonding coupling and the high symmetry of the perovskite structure for the superior photovoltaic properties of lead halide perovskite absorbers.[21–24] Therefore, lone-pair $Bi^{3+}/Sb^{3+}$ trivalent cations have been employed to replace Pb, resulting in $A_3(Bi^{3+}/Sb^{3+})_2X_9$ low-dimensional perovskite and non-perovskite (also called perovskite alternative) compounds, which show low electronic and structural dimensionalities and associated large bandgaps (> 2 eV), heavy carrier effective masses, detrimental defect properties, and thus poor photovoltaic performances.[11,21,25] To overcome the low structural dimensionality issue, trivalent $Bi^{3+}/Sb^{3+}$ has been combined with monovalent cations on the *B*-sites of halide perovskites to form $A_2B^+(Bi^{3+}/Sb^{3+})X_6$ ($B^+$ = monovalent metal) double perovskites. Since they adopt a crystal structure (cubic, space group *Fm*−3*m*) with a three-dimensional extended array of corner-sharing metal halide octahedra, a preferred characteristic for promising photovoltaic absorbers, metal halide double perovskites have received extensive attention. A recent report has systematically studied the stability and electronic properties of $A_2B^+(Bi^{3+}/Sb^{3+})X_6$ systems.[26] Experimentally, some metal halide double perovskites, e.g., $Cs_2NaB^{3+}Cl_6$ ($B^{3+}$ = Sc,[27] Y,[27] In,[27] Tl,[27] Sb[27] and Bi[28]), $Cs_2KB^{3+}Cl_6$ ($B^{3+}$ = Sc,[29] In[30]), $(MA)_2KBiCl_6$,[31] $Cs_2AgB^{3+}Cl_6$ ($B^{3+}$ = In,[32] Bi[33]), $Cs_2TlTlCl_6$[34] and $(MA)_2TlBiCl_6$[35] have been synthesized, among which, $Cs_2AgBiBr_6$ and $Cs_2AgBiCl_6$, have been proposed as candidates for photovoltaic application.[15,33,36,37] These systems also show some promising photovoltaic properties, including long carrier recombination lifetime[15] and good stability against air/moisture,[33,37] however, they possess indirect bandgaps,[15,33,35–40] an undesirable feature for efficient thin-film solar cell applications. Therefore, the search for double perovskites with direct bandgaps has been extended beyond the $Bi^{3+}/Sb^{3+}$ based systems to $A_2B^+B^{3+}X_6$ double perovskites ($B^+/B^{3+}$ = monovalent/trivalent metal). Recent reports have shown, for example, that $Cs_2AgInCl_6$ exhibits a direct bandgap.[32]

   The crystal structures of metal halide perovskites and double perovskites possess inversion symmetry, which provides the possibility of inducing parity-forbidden transitions between conduction/valence band edges, and which can in turn seriously affect materials optical absorption properties. For example, parity-forbidden transitions have been reported in other semiconductors with inversion symmetry, including $CuM^{3+}O_2$ ($M^{3+}$ = Al, Ga, In),[41] $In_2O_3$,[42] $Tl_2O_3$,[43] and $SnO_2$,[44] and were used to explain their anomalously large optical bandgaps. Inversion symmetry-induced dipole forbidden transitions are particularly undesirable for thin-film solar cell absorber application, because of the inefficient absorption of photons with energies close to the bandgap values. Though metal halide perovskites and double perovskites possess inversion symmetry, the effects of possible parity-forbidden transitions have not been systematically investigated, except for the explanation of the reported bandgap variation among groups studying $Cs_2SnI_6$.[16]

In this work, we examine the nature (direct vs indirect) of bandgaps of Pb-free metal halide perovskites and double perovskites and the effects of inversion symmetry-induced parity-forbidden transitions for the double perovskites with direct bandgaps, using density-functional theory (DFT) calculations. We show that Pb-free metal halide perovskites beyond $B^{2+}$ = Sn and Ge have indirect bandgaps and, therefore, are expected to exhibit poor optical absorptions. Among all nine possible types of Pb-free metal halide double perovskites $A_2B^+B^{3+}X_6$ ($B^+$ = monovalent metal, $B^{3+}$ = trivalent metal), distinguished by the groups from the periodic table that the $B^+$ and $B^{3+}$ metals come from, six have direct bandgaps. Of these six types, only one type, i.e., $B^+$ = In, Tl and $B^{3+}$ = Sb, Bi, shows strong transitions between conduction and valence band edges and small effective masses for carriers. All others show either symmetry-induced parity-forbidden or weak transitions, making them less suitable for thin-film solar cell applications. The parity-forbidden transition can explain very well the mystery observed with the new lead-free halide double perovskite, $Cs_2AgInCl_6$—i.e., while the synthesized powders exhibit white coloration with an experimentally measured optical bandgap of 3.3 eV, the photoluminescence (PL) emission energy is only 2.0 eV.[32] Our results suggest that inversion symmetry-induced parity-forbidden transitions must be considered when designing new double perovskites for optoelectronic application.

We first show the results of Pb halide perovskites, which will serve as a reference for comparing with the results of Pb-free perovskites and double-perovskites. For the purpose of efficient computation resource use, we consider Cs as the A cation. Figure 1(a) shows the PBE calculated band structure of CsPbI$_3$ with the α perovskite phase, which is stable at elevated temperature.[45] The conduction band minimum (CBM) is mainly derived from Pb 6p orbitals (marked in red) and exhibits an odd parity (R$_4^-$). The valence band maximum (VBM) consists of Pb 6s and I 5p and exhibits an even parity (R$_1^+$). Therefore, CsPbI$_3$ perovskite has a direct bandgap at the R point and exhibits no parity-forbidden (even to even or odd to odd) transitions.

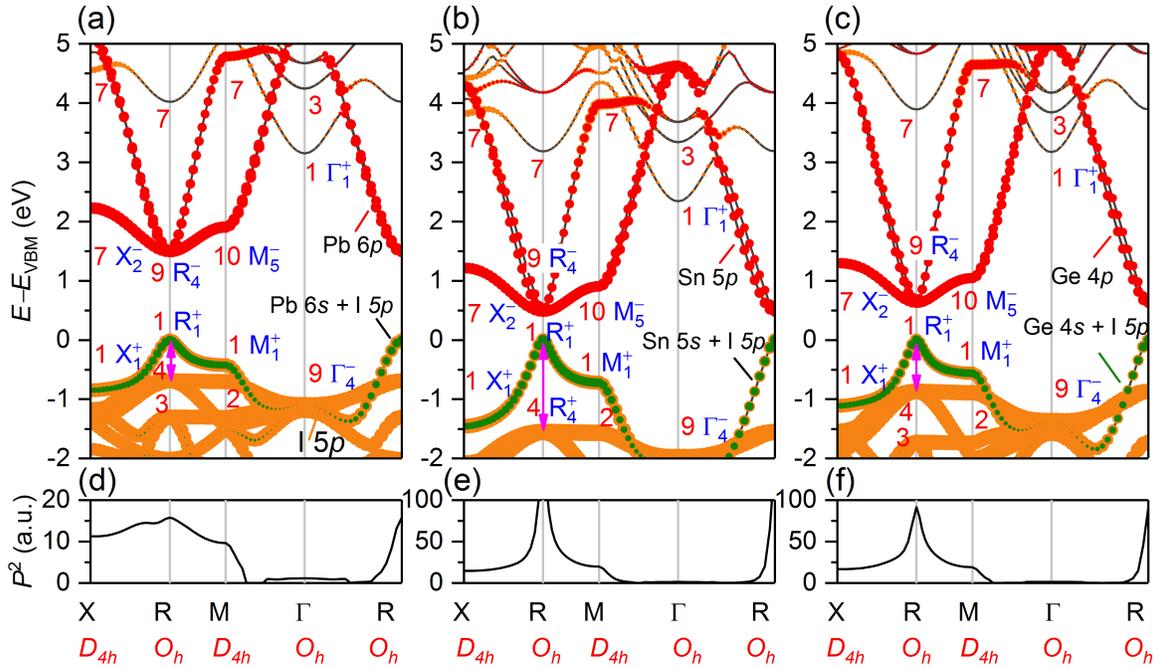

**Figure 1.** PBE calculated (a-c) band structures and (d-f) corresponding matrix elements for cubic CsB$^{2+}$I$_3$ perovskites: (a,d) CsPbI$_3$, (b,e) CsSnI$_3$, and (c,f) CsGeI$_3$. The states colored in green, red, and orange are for cation B$^{2+}$ s, cation B$^{2+}$ p and I p states, respectively. The red numbers indicate the order of Mulliken symbols for irreducible representations. Blue letters are corresponding Koster notations. Red symbols below high-symmetry K points correspond to the point groups with inversion symmetry. The corresponding band structure for CsPbI$_3$ with SOC included is shown in Figure S1(a), demonstrating that SOC does not impact the symmetry analysis given here.

The high transition probabilities between the topmost valence and the lowest conduction band are revealed by the calculated sum of the squares of the dipole transition matrix elements,[16] $P^2$, at various k points as shown in Figure 1(d). Although $P^2$ values are provided in arbitrary units, the values of different structures can be directly compared. It is worth noting that the inclusion of spin-orbital coupling (SOC) only introduces a split of the Pb 6p orbital-derived conduction band and does not change the CBM parity, as detailed for the $CsPbI_3$ band structure with SOC included (Figure S1(a)). These results are consistent with the symmetry analysis of α-$MAPbI_3$ based on a tight binding model.[46] The results can also be applied to all other Pb halide perovskites. Therefore, Pb halide perovskites exhibit very high optical absorption coefficients, making them highly desirable for thin-film solar cell applications.

We next discuss the results of Pb-free $AB^{2+}X_3$ metal halide perovskites. The $B^{2+}$ cation can be a divalent cation from group 14 (e.g., Sn, Ge), group 2 (e.g., Mg, Ca, Sr, Ba) or group 12 (e.g., Zn, Cd). Transition metals are not considered because of their localized partially occupied d orbitals. Among these candidates, only Sn(2+) and Ge(2+) exhibit lone pair s orbitals, resembling Pb $6s^2$ in Pb-based perovskites. Because of this, the band structures of $CsSnI_3$ and $CsGeI_3$ perovskites, as shown in Figure 1(b) and 1(c), respectively, have very similar features as the $CsPbI_3$ band structure—i.e, the CBM is also mainly derived from the $B^{2+}$ p orbitals, lies at the R point and has an odd parity ($R_4^-$), whereas the VBM consists of $B^{2+}$ s and X p orbitals and exhibits an even parity ($R_1^+$). As a result, $CsSnI_3$ and $CsGeI_3$ perovskites also exhibit no parity-forbidden transitions and have large dipole transition matrix elements around the R point (Figure 1(e) and (f)). However, the Sn 5s orbital is higher in energy than the Pb 6s orbital. Additionally,

Sn has a smaller atomic size than Pb. Therefore, the Sn 5s-I 5p antibonding coupling is stronger than the Pb 6s-I 5p antibonding coupling. The Sn 5s-I 5p antibonding coupling state lies 1.55 eV higher than the topmost I 5p state at the R point, as indicated by the pink arrow in Figure 1(b). The corresponding value is only 0.69 eV in $CsPbI_3$ (Figure 1(a)). Therefore, the bandgap of $CsSnI_3$ is smaller than that of $CsPbI_3$. Notably, $CsGeI_3$ exhibits a larger bandgap than $CsSnI_3$. This mainly arises because the 4s energy level (-12.031 eV) of Ge sits closer to the 6s energy level of Pb (-12.401 eV) than the 5s energy level of Sn (-10.896 eV).[47] As a result, the Ge 4s-I 5p antibonding coupling is considerably weaker than that of Sn 5s-I 5p. The Ge 4s-I 5p antibonding coupling state is about 0.87 eV higher than the topmost I 5p state at the R point (marked by the pink arrow on Figure 1(c)), which is significantly smaller than the value of 1.55 eV for $CsSnI_3$ (Figure 1(b)). Detailed analysis of the VBM components at the R point reveals that the ratios between the contributions from the $B^{2+}$ s states and the I 5p states are 0.75, 0.63 and 0.5 for $CsSnI_3$, $CsGeI_3$ and $CsPbI_3$, respectively.

If Pb is replaced by a divalent element that has no lone-pair s orbitals, such as group 2 (Mg, Ca, Sr, Ba) or group 12 (Zn, Cd), the CBM no longer derives from the $B^{2+}$ p orbital, leading to indirect bandgaps, an undesirable situation for thin-film solar cell application. As an example, Figure 2(a) shows the PBE calculated band structure of $CsZnCl_3$, a representative for $B^{2+}$ = group 12 elements. The CBM mainly derives from the Zn s orbital, which lies at the $\Gamma$ point and exhibits a $\Gamma_1^+$ even parity. The VBM comes from the Cl 3p orbital and lies at the R point. As a result, $CsZnCl_3$ perovskite exhibits an indirect bandgap. Figure 2(b) shows the PBE calculated band structure of $CsCaCl_3$, serving as an example for the case of $B^{2+}$ = group 2 elements. In this case, the CBM derives from Ca 3d orbitals and also lies at the $\Gamma$ point and exhibits a $\Gamma_1^+$ even parity. The VBM still comprises Cl p states and lies at the R point, resulting in an indirect bandgap. Substituting Pb by other group 2 and group 12 elements does not change the bandgap nature. Therefore, Pb-free perovskites are not ideal absorbers for thin-film solar cell application if Pb is replaced by elements other than Sn(2+) or Ge(2+). It should be noted that Sn(2+) and Ge(2+) can be easily oxidized into the more stable states of Sn(4+) and Ge(4+). As reported in literature, Sn(2+) and Ge(2+) halide perovskites suffer from instability issues.[8,48]

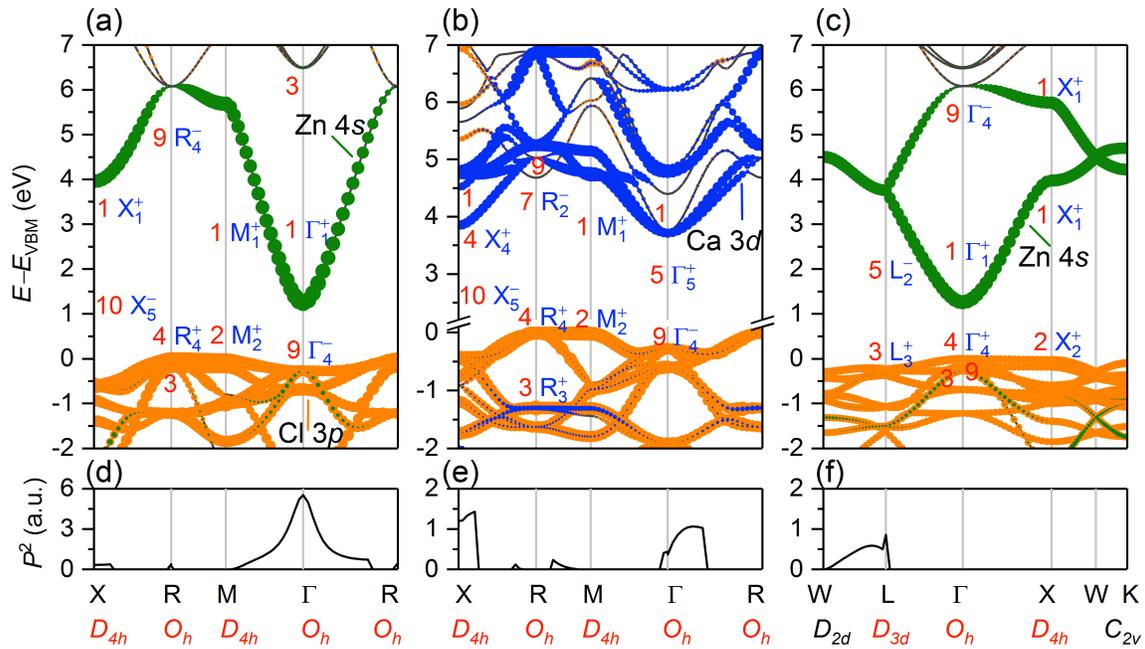

**Figure 2.** PBE calculated band structures and corresponding transition matrix elements (in arbitrary unit) for (a,d) $CsZnCl_3$ and (b,d) $CsCaCl_3$ perovskite, and (c,f) hypothetical cubic $Cs_2ZnZnCl_6$ double perovskite. The orange color corresponds to Cl 3p states, the green color corresponds to Zn 4s states and the blue color corresponds to Ca 3d states. Black/red symbols below high-symmetry K points correspond to the point groups without/with inversion symmetry.

The above results show that some metal halide perovskites do not exhibit parity-forbidden transitions between band edges, even though they possess inversion symmetry. However, as seen in Figure 2(a) and (b), the CBM and VBM have the same parity when Pb is replaced by a group 12 or group 2 element. When the CBM and VBM are located at different k points, this does not induce parity-forbidden transitions between the band edges. However, with reduced symmetry, for example in the case of double perovskites, the CBM and VBM could fold to the same k point, leading to parity-forbidden transitions between CBM and VBM. As an example, we calculated the band structure of a hypothetical cubic $Cs_2ZnZnCl_6$ double perovskite. As shown in Figure 2(c), the CBM still derives from the Zn s orbital and the VBM derives from Cl p orbital; however, now they both lie at the $\Gamma$ point, showing $\Gamma_1^+$ and $\Gamma_4^+$ even parity, respectively, and exhibiting a parity-forbidden transition between the CBM and VBM. As shown in Figure 2(f), the calculated dipole element matrix amplitude is zero between CBM and VBM at the $\Gamma$ point. Therefore, inversion symmetry-induced parity-forbidden transitions may appear in metal halide double perovskites. For comparison, the calculated dipole transition matrix elements between band edges for $CsZnCl_3$ and $CsCaCl_3$ are shown in Figure 2(d) and (e), respectively.

Table 1 shows all possible types of $A_2B^+B^{3+}X_6$ double perovskites and their corresponding bandgap nature, calculated at the PBE level. The PBE calculated lattice constants and bandgap values for considered double perovskites are given in Table S1. Among all nine possible types of $A_2B^+B^{3+}X_6$ double perovskites, three exhibit indirect or mostly-indirect bandgaps—i.e., when $B^+$ and $B^{3+}$ have the following combinations: 1) $B^+$= Cu, Ag, Au (group 11) plus $B^{3+}$ = Sc, Y (group 3) or Sb or Bi (group 15) and 2) $B^+$ = In, Tl (group 13) plus $B^{3+}$ = Al, Ga, In, Tl (group 13). The PBE calculated differences between direct and indirect bandgaps appear in Figure S2. Since an indirect bandgap is less suitable for thin-film solar cell application, these types of double perovskites are not discussed further in this study.

**Table 1.** Possible combinations for $A_2B^+B^{3+}X_6$ double perovskite.

| $B^{3+}$ / $B^+$ | | Group 3 | Group 13 | Group 15 |
|---|---|---|---|---|
| | | Sc, Y | Al, Ga, In(3+), Tl(3+) | Sb(3+), Bi (3+) |
| Group 1 | Na, K, Rb, Cs | Direct (*All forbidden*) | Direct (*All forbidden*) | Mostly Direct (*Weak transition*) |
| Group 11 | Cu, Ag, Au | Indirect | Direct (*partial forbidden*) | Indirect |
| Group 13 | In(1+), Tl(1+) | Direct (*Weak transition*) | Indirect | Direct (*Strong transition*) |

We next consider the parity of the conduction and valence bands near the bandgap position, for the six types of double perovskites exhibiting direct or mostly direct bandgaps. Four scenarios arise: (i) scenario S1: all k points forbidden. Two of the six types, i.e., $B^+$ = group 1 plus $B^{3+}$ = group 13, and $B^+$ = group 1 plus $B^{3+}$ = group 3, exhibit parity-forbidden transitions between band edges at all k points; (ii) scenario S2: partial k point forbidden. One of the six types, i.e., $B^+$ = group 11 plus $B^{3+}$ = group 13, exhibit parity-forbidden only between CBM and VBM, but with allowed transitions between band edges at other k points; (iii) scenario S3: allowed weak transitions. Two of the six types, i.e., $B^+$ = group 1 plus $B^{3+}$ = group 15, and $B^+$ = group 13 (In and Tl) plus $B^{3+}$ = group 3, exhibit allowed but weak transitions between band edges; and (iv) scenario S4: allowed strong transitions. Only one of the six types, i.e., $B^+$ = group 13 (In, Tl) plus $B^{3+}$ = group 15 (Sb, Bi), exhibit allowed and strong transitions between band edges. From the optical absorption and carrier effective mass point of view, the last type of double perovskites is most desirable for thin-film solar cell application.

Figure 3(a) shows the PBE calculated band structure of $Cs_2KInCl_6$, representing the $B^+$ = group 1 plus $B^{3+}$ = group 13 type in the scenario S1. The corresponding partial density of states (PDOS) of this double perovskite is shown in Figure S3(a). We use A = Cs and X = Cl, as an example here, but expect the results to hold for other members of the double perovskite family as well. It is worth noting that the focus of this study is on the optical absorption properties of a broad range of possible cubic double perovskites. Thermodynamic stability and bandgap values are also important, but are beyond the scope of this discussion. The orbital contributions to the states near the band gap appear in Figure 3(a). The details of the numbers, symmetry elements and Koster notations used in this paper are listed in Table S2. The lowest conduction band state derives from the In 5s and Cl 3p orbitals. The topmost valence band state derives from Cl 3p orbitals. Both the CBM and VBM lie at the $\Gamma$ point and have even parity ($\Gamma^+$). The CB and VB edge states at all other k points also have the same parity. Therefore, there are parity-forbidden transitions between band edges at all k points. As a result, the calculated transition matrix amplitudes are zero for transitions between CB and VB edges at all k points, as shown in Figure 3(c). Figure 3(b) shows the PBE calculated band structure of $Cs_2KYCl_6$, a representative double perovskite for the $B^+$ = group 1 plus $B^{3+}$ = group 3 type in scenario S1. The PDOS of this double perovskite appears in Figure S3(b). The topmost valence state derives from Cl p orbitals, while the lowest conduction state is mostly derived from the Y d orbital, except for the $\Gamma$ point, which shows the same even parity. It is seen that the CB and VB edges at all k points exhibit the same parity, introducing parity-forbidden transitions between band edges, consistent with the calculated zero transition matrix amplitudes shown in Figure 3(d). Such parity-forbidden transitions at all k points will lead to very poor optical absorptions of photons with energies close to the bandgap values, consistent with the calculated optical absorption coefficients (Figure S4(a, b)). Therefore, these two types of double perovskites are highly undesirable for thin-film solar cell application.

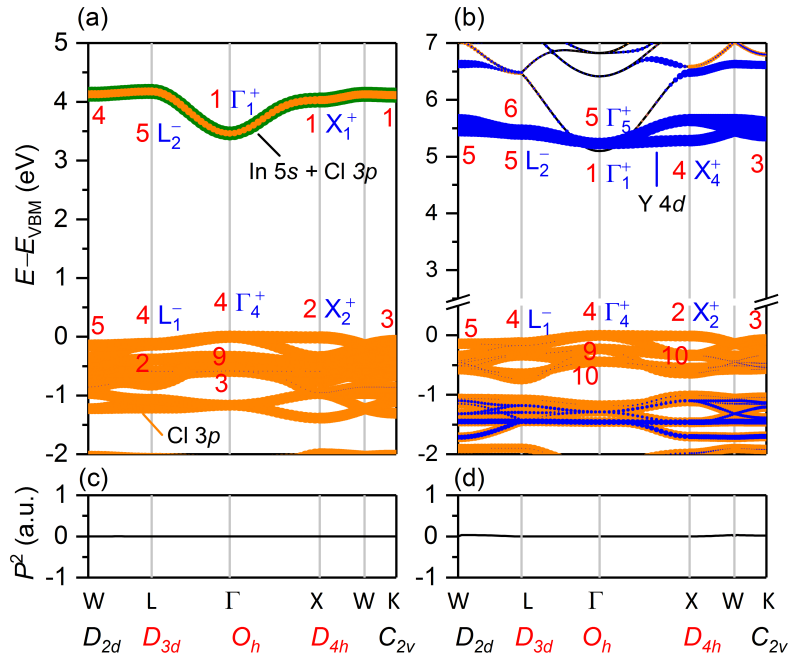

**Figure 3.** PBE calculated band structures and transition matrix elements for (a, c) $Cs_2KInCl_6$ and (b, d) $Cs_2KYCl_6$. The orange color corresponds to Cl 3p states, the green color corresponds to In 5s states and the blue color corresponds to Y 4d states. Black/red symbols below high-symmetry K points correspond to the point groups without/with inversion symmetry.

Figure 4(a) shows the PBE calculated band structure of $Cs_2AgInCl_6$, a representative of the type of $B^+$ = group 11 plus $B^{3+}$ = group 13 double perovskite in the scenario S2. Both the VBM and CBM lie at the $\Gamma$ point providing a direct bandgap. The PDOS of this double perovskite is shown in Figure S3(c). The VBM mainly derives from Ag 4d and Cl 3p orbitals. The flat band nature of the VBM along the Γ-X direction results from the disconnected nature of the Ag d orbitals along the [11$\bar{1}$] direction. The CBM mainly derives from the delocalized In 5$s$ states, which leads to a dispersive conduction band with a bandwidth of ~2.5 eV. The Ag 5s states have a much higher energy position than the In 5s states, corresponding to the CBM+1 state at the L point. Owing to the In 5s state and inversion symmetry, the CBM has a $A_{1g}$ ($\Gamma_1^+$) representation at the Γ point, as indicated in Figure 4(a). At the Γ point, the $E_g$ state of the Ag d orbitals has the highest energy. As a result, the VBM has an $E_g$ ($\Gamma_3^+$) representation at the Γ point, as indicated in Figure 4(a). Since VBM and CBM have the same even parity, a parity-forbidden transition from VBM to CBM at the Γ point should occur. However, the parity of the CB edge changes, while that of the VB edge does not, at k points away from the Γ point. Therefore, the calculated dipole transition matrix amplitude increases as the k point changes from Γ to L point, as shown in Figure 4(b). This type of double perovskite therefore has weak optical absorption coefficients for photons with energies close to the bandgap—i.e., still not an ideal condition for thin-film solar cell application (see calculated optical absorption coefficients in later discussions).

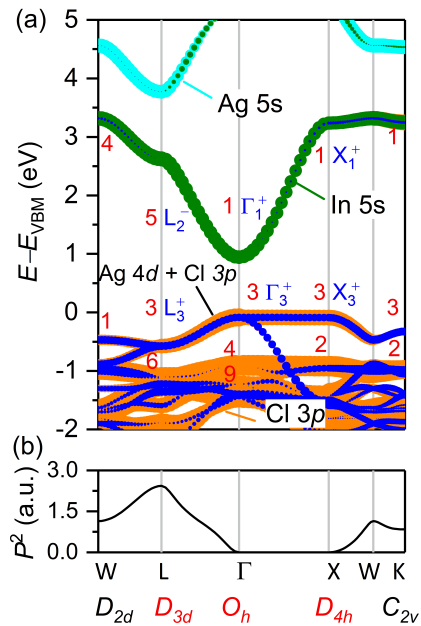

**Figure 4.** PBE calculated (a) band structure and (b) transition matrix elements for $Cs_2AgInCl_6$. The orange color corresponds to Cl 3p states, the green color corresponds to In 5s states and the blue/cyan color corresponds to Ag 4d/5s states.

Figure 5(a) shows the PBE calculated band structure of $Cs_2KBiCl_6$, a representative example of the type of $B^+$ = group 1 plus $B^{3+}$ = group 15 system in scenario S3. In this case, both the

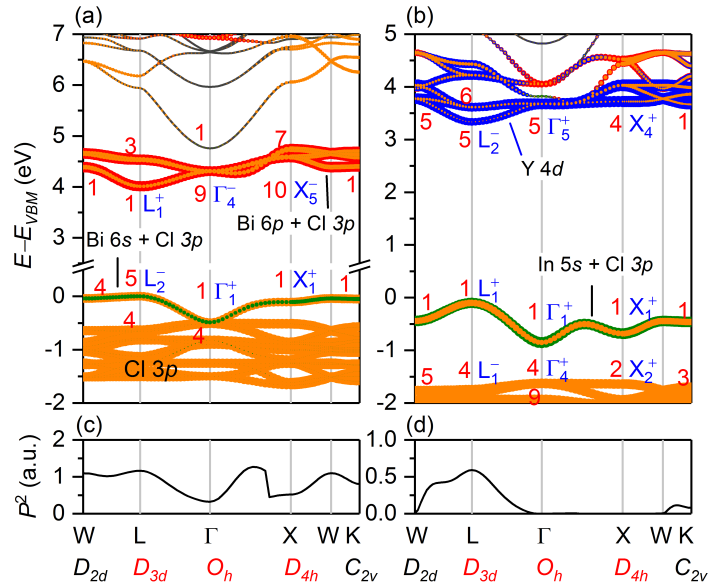

**Figure 5.** PBE calculated band structures and transition matrix elements for (a,c) $Cs_2KBiCl_6$ and (b, d) $Cs_2InYCl_6$. The orange color corresponds to Cl 3p states, the green color corresponds to Bi 6s or In 5s states, the red color corresponds to the Bi 6p states and the blue color corresponds to Y 4d states.

CBM and VBM lie at the L point. The PDOS of this double perovskite appears in Figure S3(d). The lowest conduction band primarily derives from Bi 6p and Cl 3p orbitals. The VBM also lies at the L point and mainly comes from Bi 6s and Cl 3p states. Although the bandgap is direct and there are no parity-forbidden transitions between band edges, the calculated dipole matrix amplitudes are very weak, as shown in as shown in Figure 5(c). The inclusion of SOC (Figure S1(b)) does not change the conclusion on the nature of bandgap and parity. It is noted that the Bi 6s and Cl 3p derived top VB band is rather flat, indicating a high joint density of states (JDOS) between the VB and CB band edges. According to Fermi's Golden rule, the optical absorption of a semiconductor at a specified photon energy directly relates to the dipole transition matrix elements (weak in this case) and the JDOS (strong in this case).[22] As a result, the PBE calculated absorption coefficients are moderate in value for photons with energies close to bandgap (Figure S4(c)). However, the flat CBM and VBM indicate heavy effective masses for electrons and holes, which is not preferred for solar cell application. Figure 5(b) shows the PBE calculated band structure of $Cs_2InYCl_6$, which represents the $B^+$ = group 13 (In, Tl) plus $B^{3+}$ = group 3 (Sc, Y) type of double perovskite in scenario S3. From the PDOS of this double perovskite (Figure S3(e)), the topmost VB state mainly derives from In 5s and Cl 3p orbitals, while the VBM lies at the L point. The lowest conduction band is mainly derived from Y d orbitals and the CBM also lies at the L point. While the VBM has a $L_1^+$ representation, the CBM has a $L_2^-$ representation, showing no parity-forbidden transitions between VBM and CBM. Though the calculated transition matrix amplitudes shown in Figure 5(d) are very weak between band edges at all k points, the large JDOS make the PBE calculated absorption coefficients more moderate in scale (Figure S4 (d)). However, the large bandgap renders $Cs_2InYCl_6$ not very suitable for thin-film solar cell applications.

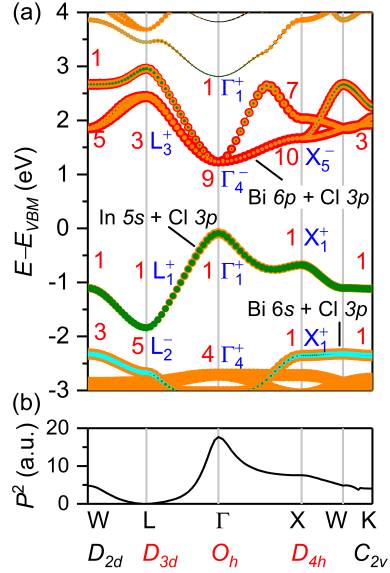

**Figure 6.** PBE calculated (a) band structures and (b) transition matrix elements for $Cs_2InBiCl_6$. The orange, green, red, and cyan colors correspond to Cl 3p, In 5s, Bi 6p and Bi 6s states, respectively.

Figure 6(a) shows the calculated band structure of $Cs_2InBiCl_6$, which represents the type of $B^+$ = group 13 (In, Tl) plus $B^{3+}$ = group 15 (Sb, Bi) double perovskite in scenario S4. The PDOS of this double perovskite (Figure S3(f)) shows that the lowest conduction band is mainly derived from the Bi 6p and Cl 3p states, while the topmost valence band is mainly derived from In 5s and Cl 3p states. Both the CBM and VBM lie at the Γ point. The CBM has a $\Gamma_4^-$ representation, while the VBM has a $\Gamma_1^+$ representation. Therefore, there is no parity-forbidden transition between CBM and VBM. The calculated transition amplitude has a large value at the Γ point (Figure 6(b)), indicating excellent optical absorption properties for the hypothetical compound

Cs$_2$InBiCl$_6$. Including SOC in the calculation (Figure S1(c)) does not change the conclusion, as compared with PBE results. The transitions between band edges are very similar to those in Pb halide perovskites. The good optical absorption (Figure S4(e)) and dispersive conduction and valence bands make this type of Pb-free double perovskite suitable for thin-film solar cell application. However, our recent combined theoretical and experimental study has shown that the In(1+)-containing halide double perovskites will be difficult to realize because of the oxidation-reduction instability of the In 1+ state in the presence of Bi 3+.[49]

Finally, we consider how parity-forbidden transitions can explain the apparent bandgap mystery observed with the new lead-free halide double perovskite, Cs$_2$AgInCl$_6$ (discussed above as an example of scenario S2 among the double perovskites). Recent reports have shown that Cs$_2$AgInCl$_6$ exhibits a direct bandgap.[32] Successful synthesis of Cs$_2$AgInCl$_6$ powders has been reported.[32] However, while the synthesized Cs$_2$AgInCl$_6$ materials exhibit white color and an experimentally-measured optical bandgap of ~3.3 eV, the PL measured from the same powder shows an emission peak at 608 nm, which corresponds to 2.0 eV.[32] The PL emission energy falls ~1.2 eV lower in energy than the optical bandgap. As shown in Figure 4(a), there are parity-forbidden transitions between CBM and VBM at the $\Gamma$ point and the transition matrix amplitude is zero at this point. Therefore, there will be no optical absorption between CBM and VBM, even for large JDOS. At k points away from the $\Gamma$ point, the symmetries of the conduction and valence band states change and the transition probability may increase. As shown in Figure 4(b), along the $\Gamma$-L line, the calculated dipole matrix amplitude slowly increases. However, all transitions along the $\Gamma$-X direction remain forbidden.

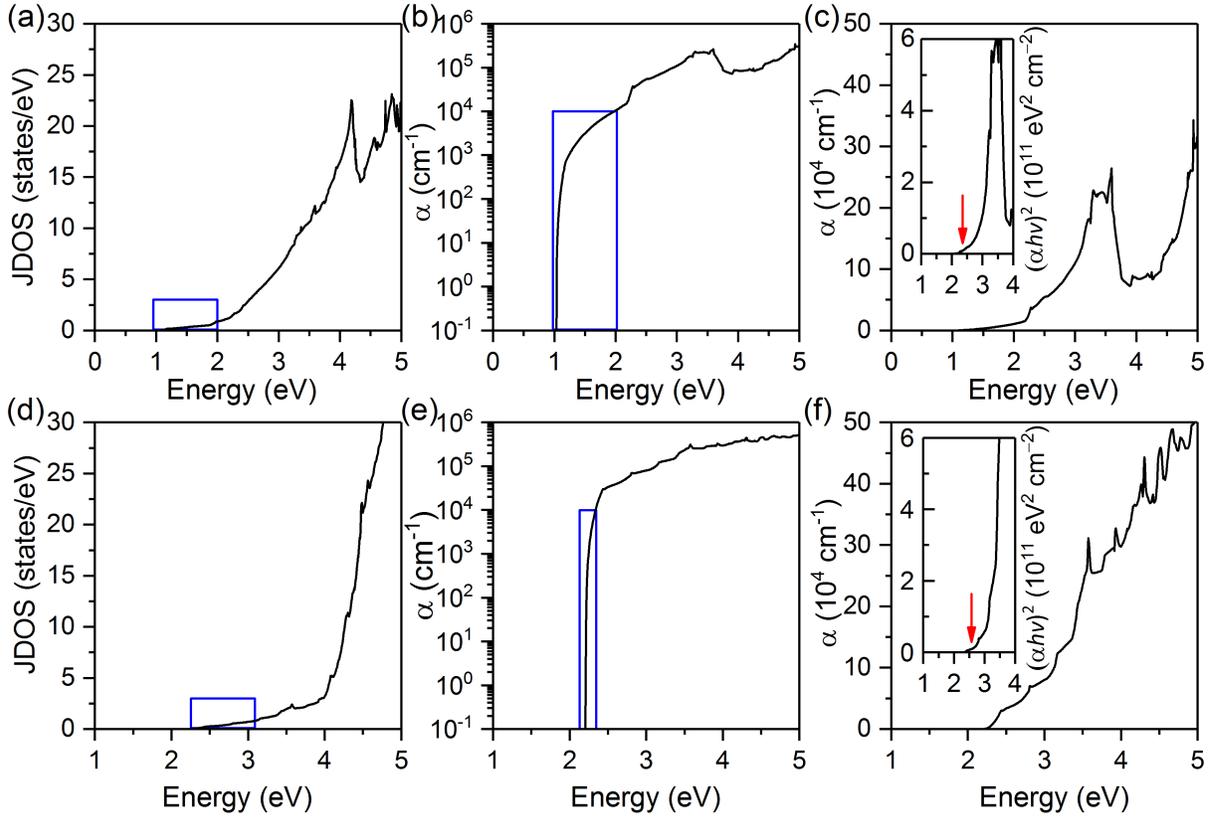

**Figure 7.** PBE calculated optical properties for $Cs_2AgInCl_6$: (a) JDOS, (b) absorption coefficient using a logarithmic scale, (c) absorption using a linear scale. Calculated optical properties for $Cs_2PbPbCl_6$: (d) JDOS, (e) absorption coefficient using a logarithmic scale and (f) absorption using a linear scale. The insets show $(\alpha h\nu)^2$ vs $h\nu$ plots to determine direct bandgap.

For comparison, we examined the JDOS values of $Cs_2AgInCl_6$ and model compound $Cs_2PbPbCl_6$ at the PBE level, as shown in Figure 7(a) and 7(d), respectively. The JDOS profiles start just slightly above their corresponding bandgap values. The JDOS values in the regions near the bandgap, marked by the blue boxes (about 1 eV wide), have rather similar values for $Cs_2AgInCl_6$ and $Cs_2PbPbCl_6$. However, their calculated optical absorption coefficient curves show rather different features. We first display the calculated absorption coefficient curves for Cs$_2$AgInCl$_6$ and Cs$_2$PbPbCl$_6$ using a logarithmic scale, as shown in Figures 7(b) and 7(e), respectively. The plots using the logarithmic scale enable easy identification of the absorption onsets, because of the rapid rise of the absorption coefficients. In both cases, the absorption onsets match well with their corresponding fundamental bandgaps extracted from their band structures. However, careful examination reveals that the two absorption coefficient curves exhibit rather different details. For Cs$_2$AgInCl$_6$, it takes up to 1 eV for the absorption coefficient to rise to $10^4$ cm$^{-1}$, as indicated by the blue box in Figure 7(b). However, for Cs$_2$PbPbCl$_6$, the energy range for the absorption coefficient to rise to the same level takes less than 0.2 eV, as shown by the blue box in Figure 7(e). Such a disparity can only be caused by differences in the transition matrix elements, since these two materials have almost the same JDOS profiles over this energy range. Notably, absorption coefficients below $10^4$ cm$^{-1}$ are regarded as weak absorption. Such weak absorption may lead to a tail in the absorption coefficient curve determined from a UV-Vis spectrum.

Weak absorption cannot easily be discerned when the absorption coefficient curves are plotted using the logarithmic scale. On the other hand, the curves plotted in a linear scale can much better reveal the effects of the weak absorption. As shown in Figures 7(c) and 7(f), while the absorption coefficient curve of Cs$_2$PbPbCl$_6$ shows a well-defined onset at about 2.3 eV, consistent with the calculated fundamental bandgap, the absorption coefficient curve of Cs$_2$AgInCl$_6$ shows an onset at about 2.1 eV and a long tail extending to about 1.0 eV, which corresponds to the PBE-calculated fundamental bandgap. The differences are seen even more clearly when the curves are plotted as $(\alpha h\nu)^2$ vs $h\nu$, which is often used to extract the direct bandgaps from experimental absorption data. As shown in the inset of Figure 7(c), the absorption begins at about 2.1 eV for the $Cs_2AgInCl_6$ double perovskite, which is about 1.0 eV above the PBE-calculated fundamental bandgap of 1.03 eV. However, for $Cs_2PbPbCl_6$ (inset in Figure 7(f)), the absorption onset is at about 2.5 eV, close to the corresponding calculated fundamental bandgap value. As discussed in the Experimental Methods section, the PBE calculation underestimates the bandgap of $Cs_2AgInCl_6$ by ~1.35 eV as compared to the HSE calculation (Figure S5). With the correction of the bandgap underestimation, the optical absorption onset of $Cs_2AgInCl_6$ should appear at ~3.45 eV, consistent with the experimental optical bandgap reported in the literature.[32] Therefore, the large difference between the absorption onset seen in an experimental $(\alpha h\nu)^2$ vs $h\nu$ plot and the bandgap calculated by HSE can be well explained by the parity-forbidden transition effect in $Cs_2AgInCl_6$. The parity-forbidden transitions only affect the absorption coefficient, but have no impact on the relaxation of the photo-excited electrons and holes, which will rapidly relax to the CBM and VBM, respectively. Therefore, the PL peak should have an energy close to the fundamental bandgap. It is expected that the PL emission may not be very strong because of the low transition matrix elements between CBM and VBM near the Γ point. Our results explain why the white powders showed PL emission at ~608 nm (2.0 eV), which is expected to reflect an orange-colored powder, if it were not for parity-forbidden transitions. The parity-forbidden transitions may also affect the carrier lifetime, making the PL lifetime longer than for a parity-allowed transition. In fact, the weak PL decay for a powder sample has shown a lifetime of up to 6μs.[32]

We have examined the bandgap nature (direct vs indirect) of Pb-free metal halide perovskites and double perovskites and the effects of inversion symmetry-induced parity-forbidden transitions for the double perovskites with direct bandgaps, using DFT calculations. We show that Pb-free metal halide perovskites beyond Sn and Ge have indirect bandgaps and, therefore, exhibit inferior optical absorptions. Among all nine possible types of Pb-free metal halide double perovskites, six have direct bandgaps. Of these six types, only one type of double perovskite, i.e., $B^+$ = In, Tl plus $B^{3+}$ = Sb, Bi, exhibits optical and electronic properties suitable for thin-film solar cell application. Parity-forbidden transitions can explain the mystery observed with the new lead-free halide double perovskite, $Cs_2AgInCl_6$—i.e., while the synthesized $Cs_2AgInCl_6$ powders exhibit white coloration with an experimentally measured optical bandgap of 3.3 eV, the photoluminescence emission energy is only 2.0 eV (consistent with the HSE-calculated bandgap). Our results suggest that inversion symmetry-induced parity-forbidden transitions must be taken into consideration when designing new double perovskites for optoelectronic application.

EXPERIMENTAL METHODS

The DFT calculations were performed using the VASP code[50,51] with the standard frozen-core projector augmented-wave (PAW)[52,53] method. For Zn, Ga, Ag, In, Tl, Pb, and Bi, the outermost d electrons were considered as valence electrons. The cutoff energy for basis functions was 400 eV. The generalized gradient approximation (GGA) of the Perdew-Burke-Ernzerh (PBE)[54] functional was used for exchange-correlation. All atoms were relaxed until the Hellmann–Feynman forces on them were below 0.01 eV/Å. For DFT calculations, the k-point meshes were chosen such that the product of the number of k points and corresponding lattice parameter are at least 40 Å and 80 Å for electronic and optical properties, respectively. SOC was considered for Tl, Pb and Bi containing systems to reveal the effects of SOC on the band structures.[55] The band symmetry and parity analysis were carried out with the QUANTUM

ESPRESSO package[56] based on GBRV PBE scalar relativistic potentials[57] and PSEUDODOJO fully relativistic potentials.[58]

For the evaluation of the band structure characteristics and analysis of orbitals and parity, only PBE was used. To validate the use of PBE and to better evaluate the bandgap of the $Cs_2AgInCl_6$ double perovskite, the hybrid functional (HSE06)[59,60] was also examined for electronic structure calculation of this compound with the PBE-relaxed structure ($a$ = 10.65 Å, exp. $a$ = 10.47 Å[32]) (Figure S5). The band structures for the PBE and HSE06 functionals have almost the same features, each exhibiting direct bandgaps. The calculated PBE and HSE06 fundamental bandgaps are 1.03 eV and 2.38 eV, respectively. The choice of PBE or HSE06 functional has minimal impact on the band dispersion and bandwidth for either the valence or conduction bands. Therefore, we used PBE calculations to analyze orbital characters, e.g. band symmetry, and optical transitions and absorptions, to save computing time.

ASSOCIATED CONTENT

AUTHOR INFORMATION

**Corresponding Author**

[*]J.W.: wang@whu.edu.cn;

[*]D.M.: david.mitzi@duke.edu;

[*]Y.Y.: yanfa.yan@utoledo.edu;

**Notes**

The authors declare no competing financial interests.

⊥Present address: Materials Research Center for Element Strategy, Tokyo Institute of Technology, Yokohama 226-8503, Japan

ACKNOWLEDGMENT

This work was funded in part by the Office of Energy Efficiency and Renewable Energy (EERE), U.S. Department of Energy, under Award Number DE-EE0006712, the Ohio Research Scholar Program, and the National Science Foundation under contract no. CHE−1230246 and DMR−1534686. This research used the resources of the National Energy Research Scientific Computing Center, which is supported by the Office of Science of the U.S. Department of Energy under Contract No. DE-AC02-05CH11231. The work at Wuhan University was supported by the National Natural Science Foundation of China (51671148, 51271134, J1210061, 11674251, 51501132, 51601132), the Hubei Provincial Natural Science Foundation of China (2016CFB446, 2016CFB155), the Fundamental Research Funds for the Central Universities, and the CERS-1-26 (CERS-China Equipment and Education Resources System).

**Supporting Information**.

Calculated lattice constant, band structure, bandgap, PDOS, band index and corresponding Mulliken symbols and Koster notation, optical absorption profile and SOC effect.

REFERENCES

(1)     Kojima, A.; Teshima, K.; Shirai, Y.; Miyasaka, T. Organometal Halide Perovskites as Visible-Light Sensitizers for Photovoltaic Cells. *J. Am. Chem. Soc.* **2009**, *131*, 6050–6051.

(2)     National Renewable Energy Laboratory (NREL). Best Research-Cell Efficiencies. *https://www.nrel.gov/pv/assets/images/efficiency-chart.png* (accessed May 10, 2017).

(3)     Zhou, H.; Chen, Q.; Li, G.; Luo, S.; Song, T. -b.; Duan, H.-S.; Hong, Z.; You, J.; Liu, Y.; Yang, Y. Interface Engineering of Highly Efficient Perovskite Solar Cells. *Science* **2014**, *345*, 542–546.

(4)     Jeon, N. J.; Noh, J. H.; Yang, W. S.; Kim, Y. C.; Ryu, S.; Seo, J.; Seok, S. Il. Compositional Engineering of Perovskite Materials for High-Performance Solar Cells. *Nature* **2015**, *517*, 476–480.

(5)     Yang, W. S.; Noh, J. H.; Jeon, N. J.; Kim, Y. C.; Ryu, S.; Seo, J.; Seok, S. I. High-Performance Photovoltaic Perovskite Layers Fabricated through Intramolecular Exchange. *Science* **2015**, *348*, 1234–1237.

(6)     Saliba, M.; Matsui, T.; Domanski, K.; Seo, J.-Y.; Ummadisingu, A.; Zakeeruddin, S. M.; Correa-Baena, J.-P.; Tress, W. R.; Abate, A.; Hagfeldt, A.; et al. Incorporation of Rubidium Cations into Perovskite Solar Cells Improves Photovoltaic Performance. *Science* **2016**, *354*, 206–209.

(7)     Hao, F.; Stoumpos, C. C.; Cao, D. H.; Chang, R. P. H.; Kanatzidis, M. G. Lead-Free Solid-State Organic–inorganic Halide Perovskite Solar Cells. *Nat. Photonics* **2014**, *8*, 489–494.

(8)     Noel, N. K.; Stranks, S. D.; Abate, A.; Wehrenfennig, C.; Guarnera, S.; Haghighirad, A.-A.; Sadhanala, A.; Eperon, G. E.; Pathak, S. K.; Johnston, M. B.; et al. Lead-Free

Organic-Inorganic Tin Halide Perovskites for Photovoltaic Applications. *Energy Environ. Sci.* **2014**, *7*, 3061–3068.

(9)     Lee, S. J.; Shin, S. S.; Kim, Y. C.; Kim, D.; Ahn, T. K.; Noh, J. H.; Seo, J.; Seok, S. Il. Fabrication of Efficient Formamidinium Tin Iodide Perovskite Solar Cells through $SnF_2$– Pyrazine Complex. *J. Am. Chem. Soc.* **2016**, *138*, 3974–3977.

(10)    Liao, W.; Zhao, D.; Yu, Y.; Grice, C. R.; Wang, C.; Cimaroli, A. J.; Schulz, P.; Meng, W.; Zhu, K.; Xiong, R.-G.; et al. Lead-Free Inverted Planar Formamidinium Tin Triiodide Perovskite Solar Cells Achieving Power Conversion Efficiencies up to 6.22%. *Adv. Mater.* **2016**, *28*, 9333–9340.

(11)    Park, B.-W.; Philippe, B.; Zhang, X.; Rensmo, H.; Boschloo, G.; Johansson, E. M. J. Bismuth Based Hybrid Perovskites $A_3Bi_2I_9$ (A: Methylammonium or Cesium) for Solar Cell Application. *Adv. Mater.* **2015**, *27*, 6806–6813.

(12)    Lyu, M.; Yun, J.-H.; Cai, M.; Jiao, Y.; Bernhardt, P. V.; Zhang, M.; Wang, Q.; Du, A.; Wang, H.; Liu, G.; et al. Organic–inorganic Bismuth (III)-Based Material: A Lead-Free, Air-Stable and Solution-Processable Light-Absorber beyond Organolead Perovskites. *Nano Res.* **2016**, *9*, 692–702.

(13)    Sun, S.; Tominaka, S.; Lee, J.-H.; Xie, F.; Bristowe, P. D.; Cheetham, A. K. Synthesis, Crystal Structure, and Properties of a Perovskite-Related Bismuth Phase, $(NH_4)_3Bi_2I_9$. *APL Mater.* **2016**, *4*, 31101.

(14)    Lehner, A. J.; Fabini, D. H.; Evans, H. A.; Hébert, C. A.; Smock, S. R.; Hu, J.; Wang, H.; Zwanziger, J. W.; Chabinyc, M. L.; Seshadri, R.; et al. Crystal and Electronic Structures of Complex Bismuth Iodides $A_3Bi_2I_9$ (A = K, Rb, Cs) Related to Perovskite: Aiding the Rational Design of Photovoltaics. *Chem. Mater.* **2015**, *27*, 7137–7148.

(15)    Slavney, A. H.; Hu, T.; Lindenberg, A. M.; Karunadasa, H. I. A Bismuth-Halide Double Perovskite with Long Carrier Recombination Lifetime for Photovoltaic Applications. *J. Am. Chem. Soc.* **2016**, *138*, 2138–2141.

(16)    Maughan, A. E.; Ganose, A. M.; Bordelon, M. M.; Miller, E. M.; Scanlon, D. O.; Neilson, J. R. Defect Tolerance to Intolerance in the Vacancy-Ordered Double Perovskite Semiconductors $Cs_2SnI_6$ and $Cs_2TeI_6$. *J. Am. Chem. Soc.* **2016**, *138*, 8453–8464.

(17)    Hong, F.; Saparov, B.; Meng, W.; Xiao, Z.; Mitzi, D. B.; Yan, Y. Viability of Lead-Free Perovskites with Mixed Chalcogen and Halogen Anions for Photovoltaic Applications. *J. Phys. Chem. C* **2016**, *120*, 6435–6441.

(18)    Saparov, B.; Sun, J.-P.; Meng, W.; Xiao, Z.; Duan, H.-S.; Gunawan, O.; Shin, D.; Hill, I. G.; Yan, Y.; Mitzi, D. B. Thin-Film Deposition and Characterization of a Sn-Deficient Perovskite Derivative $Cs_2SnI_6$. *Chem. Mater.* **2016**, *28*, 2315–2322.

(19)    Giustino, F.; Snaith, H. J. Toward Lead-Free Perovskite Solar Cells. *ACS Energy Lett.* **2016**, *1*, 1233–1240.

(20)    Xiao, Z.; Meng, W.; Saparov, B.; Duan, H.-S.; Wang, C.; Feng, C.; Liao, W. W.-Q.; Ke, W.; Zhao, D.; Wang, J.; et al. Photovoltaic Properties of Two-Dimensional $(CH_3NH_3)_2Pb(SCN)_2I_2$ Perovskite: A Combined Experimental and Density Functional Theory Study. *J. Phys. Chem. Lett.* **2016**, *7*, 1213–1218.

(21)    Xiao, Z.; Meng, W.; Wang, J.; Mitzi, D. B.; Yan, Y. Searching for Promising New Perovskite-Based Photovoltaic Absorbers: The Importance of Electronic Dimensionality. *Mater. Horiz.* **2017**, *4*, 206–216.

(22)    Yin, W.-J. J.; Shi, T.; Yan, Y. Unique Properties of Halide Perovskites as Possible Origins of the Superior Solar Cell Performance. *Adv. Mater.* **2014**, *26*, 4653–4658.

(23)    Yin, W.-J. W. W.; Yang, J. J. J.; Kang, J.; Yan, Y.; Wei, S.-H. Halide Perovskite Materials for Solar Cells: A Theoretical Review. *J. Mater. Chem. A* **2015**, *3*, 8926–8942.

(24)    Yin, W.-J.; Shi, T.; Yan, Y. Superior Photovoltaic Properties of Lead Halide Perovskites: Insights from First-Principles Theory. *J. Phys. Chem. C* **2015**, *119*, 5253–5264.

(25)    Saparov, B.; Hong, F.; Sun, J.-P.; Duan, H.-S.; Meng, W.; Cameron, S.; Hill, I. G.; Yan, Y.; Mitzi, D. B. Thin-Film Preparation and Characterization of $Cs_3Sb_2I_9$ : A Lead-Free Layered Perovskite Semiconductor. *Chem. Mater.* **2015**, *27*, 5622–5632.

(26)    Zhao, X.-G.; Yang, J.-H.; Fu, Y.; Yang, D.; Xu, Q.; Yu, L.; Wei, S.-H.; Zhang, L. Design of Lead-Free Inorganic Halide Perovskites for Solar Cells via Cation-Transmutation. *J. Am. Chem. Soc.* **2017**, *139*, 2630–2638.

(27)    Morss, L. R.; Siegal, M.; Stenger, L.; Edelstein, N. Preparation of Cubic Chloro Complex Compounds of Trivalent Metals: $Cs_2NaMCl_6$. *Inorg. Chem.* **1970**, *9*, 1771–1775.

(28)    Morrs, L. R.; Robinson, W. R. Crystal Structure of $Cs_2NaBiCl_6$. *Acta Crystallogr. Sect. B* **1972**, *28*, 653–654.

(29)    Meyer, G.; Hwu, S.-J.; Corbett, J. D. Low-Temperature Crystal Growth of $Cs_2LiLuCl_6$-II and $Cs_2KScCl_6$ under Reducing Conditions and Their Structural Refinement. *Z. Anorg. Allg. Chem.* **1986**, *535*, 208–212.

(30)    Guedira, T.; Wignacourt, J. P.; Drache, M.; Lorriaux-Rubbens, A.; Wallart, F. Phase Transition in a Chloro-Elpasolite, $Cs_2KInCl_6$. *Phase Transitions* **1988**, *13*, 81–85.

(31)    Wei, F.; Deng, Z.; Sun, S.; Xie, F.; Kieslich, G.; Evans, D. M.; Carpenter, M. A.; Bristowe, P. D.; Cheetham, A. K. The Synthesis, Structure and Electronic Properties of a Lead-Free Hybrid Inorganic–organic Double Perovskite $(MA)_2KBiCl_6$ (MA =

Methylammonium). *Mater. Horiz.* **2016**, *3*, 328–332.

(32)  Volonakis, G.; Haghighirad, A. A.; Milot, R. L.; Sio, W. H.; Filip, M. R.; Wenger, B.; Johnston, M. B.; Herz, L. M.; Snaith, H. J.; Giustino, F. $Cs_2InAgCl_6$ : A New Lead-Free Halide Double Perovskite with Direct Band Gap. *J. Phys. Chem. Lett.* **2017**, *8*, 772–778.

(33)  McClure, E. T.; Ball, M. R.; Windl, W.; Woodward, P. M. $Cs_2AgBiX_6$ (X = Br, Cl): New Visible Light Absorbing, Lead-Free Halide Perovskite Semiconductors. *Chem. Mater.* **2016**, *28*, 1348–1354.

(34)  Retuerto, M.; Emge, T.; Hadermann, J.; Stephens, P. W.; Li, M. R.; Yin, Z. P.; Croft, M.; Ignatov, A.; Zhang, S. J.; Yuan, Z.; et al. Synthesis and Properties of Charge-Ordered Thallium Halide Perovskites, $CsTl^+_{0.5}Tl^{3+}_{0.5}X_3$ (X = F or Cl): Theoretical Precursors for Superconductivity? *Chem. Mater.* **2013**, *25*, 4071–4079.

(35)  Deng, Z.; Wei, F.; Sun, S.; Kieslich, G.; Cheetham, A. K.; Bristowe, P. D. Exploring the Properties of Lead-Free Hybrid Double Perovskites Using a Combined Computational-Experimental Approach. *J. Mater. Chem. A* **2016**, *4*, 12025–12029.

(36)  Filip, M. R.; Hillman, S.; Haghighirad, A. A.; Snaith, H. J.; Giustino, F. Band Gaps of the Lead-Free Halide Double Perovskites $Cs_2BiAgCl_6$ and $Cs_2BiAgBr_6$ from Theory and Experiment. *J. Phys. Chem. Lett.* **2016**, *7*, 2579–2585.

(37)  Volonakis, G.; Filip, M. R.; Haghighirad, A. A.; Sakai, N.; Wenger, B.; Snaith, H. J.; Giustino, F. Lead-Free Halide Double Perovskites via Heterovalent Substitution of Noble Metals. *J. Phys. Chem. Lett.* **2016**, *7*, 1254–1259.

(38)  Savory, C. N.; Walsh, A.; Scanlon, D. O. Can Pb-Free Halide Double Perovskites Support High-Efficiency Solar Cells? *ACS Energy Lett.* **2016**, *1*, 949–955.

(39)  Xiao, Z.; Meng, W.; Wang, J.; Yan, Y. Thermodynamic Stability and Defect Chemistry of Bismuth-Based Lead-Free Double Perovskites. *ChemSusChem* **2016**, *9*, 2628–2633.

(40)  Feng, H.-J.; Deng, W.; Yang, K.; Huang, J.; Zeng, X. C. Double Perovskite $Cs_2BBiX_6$ (B = Ag, Cu; X = Br, Cl)/$TiO_2$ Heterojunction: An Efficient Pb-Free Perovskite Interface for Charge Extraction. *J. Phys. Chem. C* **2017**, *121*, 4471–4480.

(41)  Nie, X.; Wei, S.-H.; Zhang, S. B. Bipolar Doping and Band-Gap Anomalies in Delafossite Transparent Conductive Oxides. *Phys. Rev. Lett.* **2002**, *88*, 66405.

(42)  Walsh, A.; Da Silva, J. L. F.; Wei, S.-H.; Körber, C.; Klein, A.; Piper, L. F. J.; DeMasi, A.; Smith, K. E.; Panaccione, G.; Torelli, P.; et al. Nature of the Band Gap of $In_2O_3$ Revealed by First-Principles Calculations and X-Ray Spectroscopy. *Phys. Rev. Lett.* **2008**, *100*, 167402.

(43)  Kehoe, A. B.; Scanlon, D. O.; Watson, G. W. Nature of the Band Gap of $Tl_2O_3$. *Phys. Rev. B* **2011**, *83*, 233202.

(44)  Summitt, R.; Marley, J. A.; Borrelli, N. F. The Ultraviolet Absorption Edge of Stannic Oxide ($SnO_2$). *J. Phys. Chem. Solids* **1964**, *25*, 1465–1469.

(45)  Swarnkar, A.; Marshall, A. R.; Sanehira, E. M.; Chernomordik, B. D.; Moore, D. T.; Christians, J. A.; Chakrabarti, T.; Luther, J. M. Quantum Dot-Induced Phase Stabilization of α-$CsPbI_3$ Perovskite for High-Efficiency Photovoltaics. *Science* **2016**, *354*, 92–95.

(46)  Boyer-Richard, S.; Katan, C.; Traoré, B.; Scholz, R.; Jancu, J.-M.; Even, J. Symmetry-Based Tight Binding Modeling of Halide Perovskite Semiconductors. *J. Phys. Chem. Lett.* **2016**, *7*, 3833–3840.

(47)  Huang, L.; Lambrecht, W. R. L. Electronic Band Structure Trends of Perovskite Halides: Beyond Pb and Sn to Ge and Si. *Phys. Rev. B* **2016**, *93*, 195211.

(48)   Sun, P.-P.; Li, Q.-S.; Yang, L.-N.; Li, Z.-S. Theoretical Insights into a Potential Lead-Free Hybrid Perovskite: Substituting $Pb^{2+}$ with $Ge^{2+}$. *Nanoscale* **2016**, *8*, 1503–1512.

(49)   Xiao, Z.; Du, K.; Meng, W.; Wang, J.; Mitzi, D. B.; Yan, Y. Intrinsic Instability of $Cs_2$ In(I)M(III)$X_6$ (M = Bi, Sb; X = Halogen) Double Perovskites: A Combined Density Functional Theory and Experimental Study. *J. Am. Chem. Soc.* **2017**, *139*, 6054–6057.

(50)   Kresse, G.; Hafner, J. Ab Initio Molecular Dynamics for Liquid Metals. *Phys. Rev. B* **1993**, *47*, 558–561.

(51)   Kresse, G.; Furthmüller, J. Efficient Iterative Schemes for Ab Initio Total-Energy Calculations Using a Plane-Wave Basis Set. *Phys. Rev. B* **1996**, *54*, 11169–11186.

(52)   Blöchl, P. E.; Blochl, P. E.; Blöchl, P. E. Projector Augmented-Wave Method. *Phys. Rev. B* **1994**, *50*, 17953–17979.

(53)   Kresse, G.; Joubert, D. From Ultrasoft Pseudopotentials to the Projector Augmented-Wave Method. *Phys. Rev. B* **1999**, *59*, 1758–1775.

(54)   Perdew, J. P.; Burke, K.; Ernzerhof, M. Generalized Gradient Approximation Made Simple. *Phys. Rev. Lett.* **1996**, *77*, 3865–3868.

(55)   Even, J.; Pedesseau, L.; Jancu, J.-M.; Katan, C. Importance of Spin–Orbit Coupling in Hybrid Organic/Inorganic Perovskites for Photovoltaic Applications. *J. Phys. Chem. Lett.* **2013**, *4*, 2999–3005.

(56)   Giannozzi, P.; Baroni, S.; Bonini, N.; Calandra, M.; Car, R.; Cavazzoni, C.; Ceresoli, D.; Chiarotti, G. L.; Cococcioni, M.; Dabo, I.; et al. QUANTUM ESPRESSO: A Modular and Open-Source Software Project for Quantum Simulations of Materials. *J. Phys. Condens. Matter* **2009**, *21*, 395502.

(57)   Garrity, K. F.; Bennett, J. W.; Rabe, K. M.; Vanderbilt, D. Pseudopotentials for High-

Throughput DFT Calculations. *Comput*. *Mater*. *Sci*. **2014**, *81*, 446–452.

(58)     Lejaeghere, K.; Bihlmayer, G.; Bjorkman, T.; Blaha, P.; Blugel, S.; Blum, V.; Caliste, D.; Castelli, I. E.; Clark, S. J.; Dal Corso, A.; et al. Reproducibility in Density Functional Theory Calculations of Solids. *Science* **2016**, *351*, 1415.

(59)     Heyd, J.; Scuseria, G. E.; Ernzerhof, M. Hybrid Functionals Based on a Screened Coulomb Potential. *J*. *Chem*. *Phys*. **2003**, *118*, 8207.

(60)     Krukau, A. V.; Vydrov, O. A.; Izmaylov, A. F.; Scuseria, G. E. Influence of the Exchange Screening Parameter on the Performance of Screened Hybrid Functionals. *J*. *Chem*. *Phys*. **2006**, *125*, 224106.